\documentclass[twocolumn,showkeys,showpacs,pra,byrevtex]{revtex4}%
\pdfoutput=1
\usepackage{amsfonts}
\usepackage{amsmath}
\usepackage{amssymb}
\usepackage{graphicx}%
\providecommand{\U}[1]{\protect\rule{.1in}{.1in}}

\begin{document}
\preprint{ }
\title{Linear-Optical Hyperentanglement-Assisted Quantum Error-Correcting Code}
\author{Mark M. Wilde}
\affiliation{Center for Quantum Information Science and Technology, Department of
Electrical Engineering, University of Southern California, Los Angeles,
California 90089, USA}
\author{Dmitry B. Uskov}
\affiliation{Department of Physics, Tulane University, New Orleans, Louisiana\ 70118, USA}
\affiliation{Hearne Institute for Theoretical Physics, Department of Physics and Astronomy,
Louisiana State University, Baton Rouge, Louisiana 70803, USA}

\begin{abstract}
We propose a linear-optical implementation of a hyperentanglement-assisted
quantum error-correcting code. The code is hyperentanglement-assisted because
the shared entanglement resource is a photonic state hyperentangled in
polarization and orbital angular momentum. It is possible to encode, decode,
and diagnose channel errors using linear-optical techniques. The code corrects
for polarization \textquotedblleft flip\textquotedblright\ errors and is thus
suitable only for a proof-of-principle experiment. The encoding and decoding
circuits use a Knill-Laflamme-Milburn-like scheme for transforming
polarization and orbital angular momentum photonic qubits. A numerical
optimization algorithm finds a unit-fidelity encoding circuit that requires
only three ancilla modes and has success probability equal to 0.0097.

\end{abstract}
\volumeyear{2008}
\volumenumber{ }
\issuenumber{ }
\eid{ }
\date{\today}
\received[Received text]{}

\revised[Revised text]{}

\accepted[Accepted text]{}

\published[Published text]{}

\maketitle

\section{Introduction}

Quantum error correction plays an active role in the future realization of a
quantum communication system \cite{ieee1998calderbank,thesis97gottesman}.
Several optical experiments have already implemented simple quantum error
correction routines \cite{nat2004qecoptics,o'brien:060303,pittman:052332}.

The entanglement-assisted stabilizer formalism is a recent extension of the
theory of quantum error correction that incorporates entanglement shared
between a sender and receiver \cite{science2006brun,arx2006brun}. A further
extension of this theory incorporates gauge qubits \cite{hsieh:062313} and
others give a structure appropriate for a stream of quantum information
\cite{arx2007wildeCED,arx2007wildeEAQCC,prep2008wildeGEN}. The likely
candidate for implementing an entanglement-assisted code is photonics because
the entanglement-assisted model is more appropriate for quantum communication
rather than quantum computing.

In this article, we propose a linear-optical implementation of a
hyperentanglement-assisted quantum code. Our code is
\textit{hyperentanglement-assisted} because it exploits
\textit{hyperentanglement} of two photons \cite{jmo1997kwiat}. Two photons are
hyperentangled if they have entanglement in multiple degrees of freedom such
as polarization and orbital angular momentum (OAM)
\cite{barreiro:260501,nat2007twisted}. The benefit of hyperentanglement is
that a linear-optical setup suffices to perform a complete Bell-state analysis
\cite{PhysRevA.58.R2623,schuck:190501,barbieri:042317}. Our proposal for the
hyperentanglement-assisted code relies on the recent optical realization
\cite{nat2008kwiat}\ of the superdense coding protocol
\cite{PhysRevLett.69.2881} and the close connection between
entanglement-assisted quantum error correction and superdense coding
\cite{arx2006brun}. We also employ a recent numerical optimization algorithm
\cite{prep2008uskov}\ to find an encoding circuit and a decoding circuit that
has unit fidelity, success probability equal to 0.0097, and requires only
three ancilla modes. The circuits act on both the polarization and OAM degrees
of freedom of the photonic qubits.

We structure this article as follows. The first section reviews hyperentangled
states, the single-photon polarization-OAM states, and mentions that it is
possible to distinguish the single-photon polarization-OAM states with linear
optics. We then discuss the superdense coding protocol for hyperentangled
states and highlight the connection between superdense coding and
entanglement-assisted quantum error correction. We give a brief description of
our code, its error analysis, and corrective operations. The final part of
this article discusses the numerical optimization technique for finding our
code's encoding circuit and decoding circuit.

\section{Hyperentangled States}

The standard hyperentangled state is a state of two photons simultaneously
entangled in polarization and OAM:%
\[
\frac{1}{2}\left(  \left\vert HH\right\rangle +\left\vert VV\right\rangle
\right)  \otimes\left(  \left\vert \circlearrowleft\circlearrowright
\right\rangle +\left\vert \circlearrowright\circlearrowleft\right\rangle
\right)  .
\]
The symbols $H$ and $V$ represent horizontal and vertical polarization
respectively and the symbols $\circlearrowleft$ and $\circlearrowright$
represent paraxial Laguerre-Gauss spatial modes with $+\hbar$ and $-\hbar$
respective units of OAM \cite{book2003OAM}. Changing the polarization degree
of freedom of Alice's photon in the above state according to the four standard
Pauli operators, while leaving the OAM degree of freedom unchanged, gives the
following four hyperentangled states:%
\begin{align}
\left\vert \Phi^{\pm}\right\rangle  &  \equiv\frac{1}{2}\left(  \left\vert
HH\right\rangle \pm\left\vert VV\right\rangle \right)  \otimes\left(
\left\vert \circlearrowleft\circlearrowright\right\rangle +\left\vert
\circlearrowright\circlearrowleft\right\rangle \right)  ,\nonumber\\
\left\vert \Psi^{\pm}\right\rangle  &  \equiv\frac{1}{2}\left(  \left\vert
HV\right\rangle \pm\left\vert VH\right\rangle \right)  \otimes\left(
\left\vert \circlearrowleft\circlearrowright\right\rangle +\left\vert
\circlearrowright\circlearrowleft\right\rangle \right)  .\label{eq:ebit}%
\end{align}
We can rewrite the above four states in terms of the single-photon
polarization-OAM states $\phi^{\pm},\psi^{\pm}$
\footnote{Ref.~\cite{nat2008kwiat} refers to these states as spin-orbit Bell
states, but this name is inappropriate given that the states are encoded in
polarization and orbital angular momentum and it is not clear whether these
states could violate a Bell's inequality.}:%
\begin{align*}
\left\vert \Phi^{\pm}\right\rangle  &  =\frac{1}{4}\left(  \phi_{A}^{+}%
\otimes\psi_{B}^{\pm}+\phi_{A}^{-}\otimes\psi_{B}^{\mp}+\psi_{A}^{+}%
\otimes\phi_{B}^{\pm}+\psi_{A}^{-}\otimes\phi_{B}^{\mp}\right)  ,\\
\left\vert \Psi^{\pm}\right\rangle  &  =\frac{1}{4}\left(  \pm\phi_{A}%
^{+}\otimes\phi_{B}^{\pm}\mp\phi_{A}^{-}\otimes\phi_{B}^{\mp}\pm\psi_{A}%
^{+}\otimes\psi_{B}^{\pm}\mp\psi_{A}^{-}\otimes\psi_{B}^{\mp}\right)  ,
\end{align*}
where $A$ and $B$ label the first and second respective photons and the
single-photon polarization-OAM states $\phi^{\pm}$ and $\psi^{\pm}$\ are as
follows:%
\[
\phi^{\pm}\equiv\frac{\left\vert H\circlearrowleft\right\rangle \pm\left\vert
V\circlearrowright\right\rangle }{\sqrt{2}},\ \ \ \ \psi^{\pm}\equiv
\frac{\left\vert H\circlearrowright\right\rangle \pm\left\vert
V\circlearrowleft\right\rangle }{\sqrt{2}}.
\]
The above \textquotedblleft quad-rail\textquotedblright\ basis states
$\left\vert H\circlearrowleft\right\rangle $, $\left\vert H\circlearrowright
\right\rangle $, $\left\vert V\circlearrowleft\right\rangle $, $\left\vert
V\circlearrowright\right\rangle $ are four-mode single-photon states defined
in terms of the Fock basis:%
\begin{align}
\left\vert H\circlearrowleft\right\rangle  &  \equiv\left\vert 1\right\rangle
_{H\circlearrowleft}\left\vert 0\right\rangle _{H\circlearrowright}\left\vert
0\right\rangle _{V\circlearrowleft}\left\vert 0\right\rangle
_{V\circlearrowright},\nonumber\\
\left\vert H\circlearrowright\right\rangle  &  \equiv\left\vert 0\right\rangle
_{H\circlearrowleft}\left\vert 1\right\rangle _{H\circlearrowright}\left\vert
0\right\rangle _{V\circlearrowleft}\left\vert 0\right\rangle
_{V\circlearrowright},\nonumber\\
\left\vert V\circlearrowleft\right\rangle  &  \equiv\left\vert 0\right\rangle
_{H\circlearrowleft}\left\vert 0\right\rangle _{H\circlearrowright}\left\vert
1\right\rangle _{V\circlearrowleft}\left\vert 0\right\rangle
_{V\circlearrowright},\nonumber\\
\left\vert V\circlearrowright\right\rangle  &  \equiv\left\vert 0\right\rangle
_{H\circlearrowleft}\left\vert 0\right\rangle _{H\circlearrowright}\left\vert
0\right\rangle _{V\circlearrowleft}\left\vert 1\right\rangle
_{V\circlearrowright}.\label{eq:quad-rail}%
\end{align}
Hyperentangled states are useful because a \textit{linear-optical} analyzer
distinguishes the single-photon polarization-OAM states $\phi^{\pm},\psi^{\pm
}$ and thus distinguishes the hyperentangled states $\left\vert \Phi^{\pm
}\right\rangle ,\left\vert \Psi^{\pm}\right\rangle $ as well
\cite{PhysRevA.58.R2623,schuck:190501,barbieri:042317}.

\section{Superdense Coding and Entanglement-assisted Quantum Error Correction}

We briefly describe the superdense coding protocol for polarization encoding
and hyperentangled states \cite{nat2008kwiat}. A sender Alice and a receiver
Bob share a hyperentangled state $\left\vert \Phi^{+}\right\rangle ^{AB}$.
Alice encodes one of four classical messages (two classical bits)\ by applying
one of four transformations to her half of $\left\vert \Phi^{+}\right\rangle
^{AB}$: (1) the identity, (2) $\left\vert V\right\rangle \rightarrow
-\left\vert V\right\rangle $, (3)\ $\left\vert H\right\rangle \leftrightarrow
\left\vert V\right\rangle $, or (4) both $\left\vert V\right\rangle
\rightarrow-\left\vert V\right\rangle $ and $\left\vert H\right\rangle
\leftrightarrow\left\vert V\right\rangle $. Let $Z$ denote the second
operation and let $X$ denote the third operation. The result is to transform
the original state $\left\vert \Phi^{+}\right\rangle ^{AB}$ to one of the four
states $\left\vert \Phi^{\pm}\right\rangle ^{AB},\left\vert \Psi^{\pm
}\right\rangle ^{AB}$. She then sends her half of the encoded $\left\vert
\Phi^{+}\right\rangle ^{AB}$ to Bob. Bob performs a single-photon
polarization-OAM state analysis in the basis $\phi^{\pm},\psi^{\pm}$\ on each
of the systems $A$ and $B$\ to distinguish the message Alice transmitted.

In the above analysis, it is important to stress that the dense coding
transformations affect only the polarization degree of freedom of the
hyperentangled state $\left\vert \Phi^{+}\right\rangle ^{AB}$. The classical
information resides in a four-dimensional subspace of the 16-dimensional
Hilbert space. The extra dimensions help in single-photon polarization-OAM
state analysis in order to distinguish the classical messages.

Ref.~\cite{arx2006brun} discusses the close relationship between superdense
coding and entanglement-assisted quantum error correction. In superdense
coding, one exploits the classical bits encoded in a Bell state so that Alice
can transmit classical information to Bob. In entanglement-assisted error
correction, one exploits the encoded classical bits for use as error
syndromes. Another way of thinking about this latter scenario is that Eve, the
environment, is superdense coding messages\ (errors) into the Bell states. Bob
can determine the errors that Eve introduces by measuring in the Bell basis.

\section{Operation of the Code}

\begin{figure}
[ptb]
\begin{center}
\includegraphics[
natheight=3.039800in,
natwidth=8.460500in,
height=0.934in,
width=3.3338in
]%
{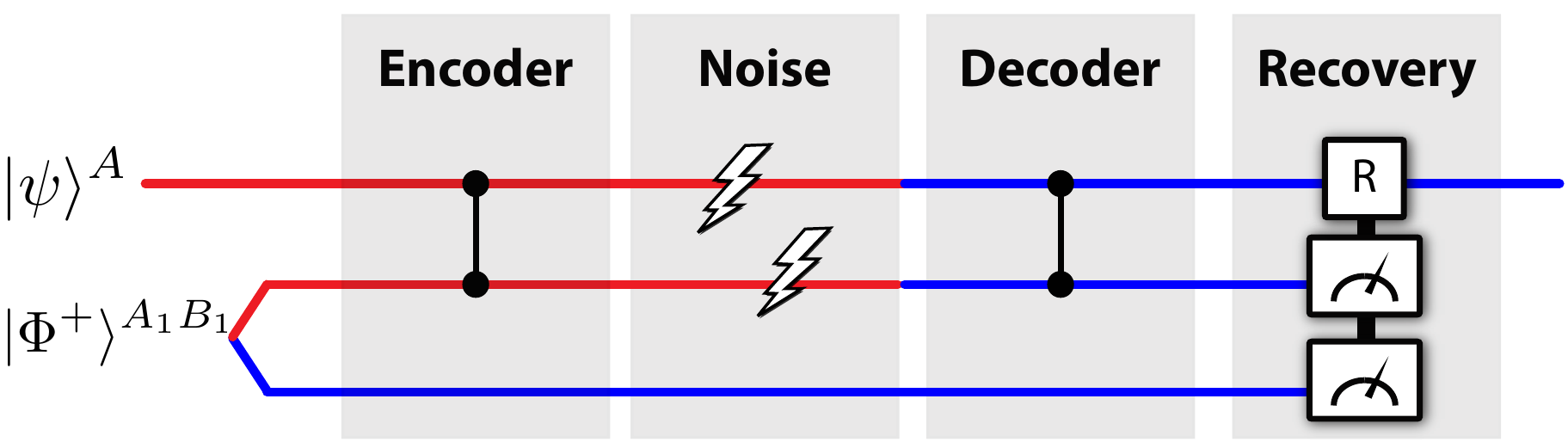}%
\caption{(Color online) The operation of the hyperentanglement-assisted
quantum code. Red qubits (those labeled as $A$ and $A_{1}$) belong to the
sender Alice and blue qubits (the one labeled as $B_{1}$) belong to the
receiver Bob (though all qubits belong to Bob after the noisy channel). Alice
and Bob share a hyperentangled state $\left\vert \Phi^{+}\right\rangle
^{A_{1}B_{1}}$ prior to quantum communication. Alice uses the hyperentangled
state to aid in encoding an information photon in the state $\left\vert
\psi\right\rangle ^{A}$. Her encoding circuit consists of one controlled-phase
gate. She sends her photons over a noisy polarization-error quantum channel.
Bob receives the photons, performs a decoding circuit, and performs two
single-photon polarization-OAM\ analyses in the basis $\phi^{\pm},\psi^{\pm}$
on the systems $A_{1}$ and $B_{1}$ to determine the error syndrome. Bob
finally performs a recovery operation to obtain the information photon
$\left\vert \psi\right\rangle ^{A}$ that Alice first sent.}%
\label{fig:circuit}%
\end{center}
\end{figure}

The operation of our code begins with an initial, unencoded state consisting
of one information photon and one hyperentangled state:%
\begin{equation}
\left\vert \psi\right\rangle ^{A}\left\vert \Phi^{+}\right\rangle ^{A_{1}%
B_{1}}.\label{eq:unencoded-state}%
\end{equation}
The information photon is as follows: $\left\vert \psi\right\rangle ^{A}%
\equiv\alpha\left\vert H\right\rangle ^{A}+\beta\left\vert V\right\rangle
^{A}$. The sender Alice possesses photons $A$ and $A_{1}$ and the receiver Bob
possesses photon $B_{1}$. The entanglement-assisted communication paradigm
assumes that Alice and Bob share the hyperentangled state prior to quantum
communication. Figure~\ref{fig:circuit} highlights the operation of our
hyperentanglement-assisted quantum code.%

The sender Alice applies an encoding circuit consisting of one controlled-sign
gate (we discuss this gate later in more detail)\ so that the state shared
between Alice and Bob is the following unnormalized encoded state:%
\begin{align}
&  \left\vert \psi\right\rangle ^{A}\left(  \left\vert \Phi^{-}\right\rangle
^{A_{1}B_{1}}+\left\vert \Phi^{+}\right\rangle ^{A_{1}B_{1}}\right)
+\nonumber\\
&  Z\left\vert \psi\right\rangle ^{A}\left(  \left\vert \Phi^{+}\right\rangle
^{A_{1}B_{1}}-\left\vert \Phi^{-}\right\rangle ^{A_{1}B_{1}}\right)  .
\label{eq:encoded-state}%
\end{align}
She sends her photons $A$ and $A_{1}$ over a noisy polarization-error channel
(discussed below). Bob receives the photons and we relabel them as $B$ and
$B_{1}^{\prime}$ respectively. For now, suppose that the channel does not
introduce an error. Bob finally applies the decoding circuit (same as the
encoding circuit) and the resulting decoded state is as follows:%
\begin{equation}
\left\vert \psi\right\rangle ^{B}\left\vert \Phi^{+}\right\rangle
^{B_{1}^{\prime}B_{1}}, \label{eq:clean-decoded-state}%
\end{equation}
where the information photon appears in Bob's system $B$. Bob performs two
single-photon polarization-OAM analyses in the basis $\phi^{\pm},\psi^{\pm}$
on the systems $B_{1}^{\prime}$ and $B_{1}$ to diagnose the channel error. The
polarization-OAM analysis distinguishes the four states $\left\{  \left\vert
\Phi^{\pm}\right\rangle ,\left\vert \Psi^{\pm}\right\rangle \right\}  $. Bob
measures the result $\left\vert \Phi^{+}\right\rangle $ when the channel does
not introduce noise. The state $\left\vert \Phi^{+}\right\rangle \ $is a
\textit{syndrome }that determines the channel error. Bob does not need to
perform a recovery operation in this case.

\section{Error Analysis}

In general, the channel introduces errors on the photons that Alice transmits.
We assume in this article that the channel is a noisy polarization-error
channel (analogous to the classical bit-flip channel.)\ A noisy
polarization-error channel independently applies a polarization error $X$ that
flips the horizontal and vertical polarization bases. We assume that this
channel affects the polarization degree of freedom only and does not affect
the OAM degree of freedom. Although this channel may not be entirely
realistic, it provides a platform for a proof-of-principle demonstration of
the operation of an entanglement-assisted quantum code \footnote{We do have
examples of hyperentanglement-assisted codes that can correct an arbitrary
single-qubit error. But such codes either require more hyperentanglement or
increase the complexity of our numerical optimization procedure (or do both).
This increase in complexity conflicts with our main purpose in this
article---to suggest a minimal, proof-of-principle demonstration of
entanglement-assisted error correction.}.

The code protects against a single polarization error on either of the two
photons $A$ or $A_{1}$ that Alice sends. It also protects against a double
polarization error on both photons. Suppose that a polarization error occurs
on photon $A$. After Bob applies the decoding circuit, the state becomes
$X\left\vert \psi\right\rangle ^{B}\left\vert \Phi^{-}\right\rangle
^{B_{1}^{\prime}B_{1}}$. Bob measures the photons $B_{1}^{\prime}$ and $B_{1}%
$, determines they are in the state $\left\vert \Phi^{-}\right\rangle $, and
flips the polarization of photon $B$ to recover the initial information photon
$\left\vert \psi\right\rangle $. Table~\ref{tbl:syndromes} summarizes the
other cases.%

\begin{table}[tbp] \centering
\begin{tabular}
[c]{l|l|l}\hline\hline
\textbf{Error} & \textbf{Recovery} & \textbf{Syndrome}\\\hline\hline
$I$ & $I$ & \multicolumn{1}{|c}{$\Phi^{+}$}\\\hline
$X^{A}$ & $X$ & \multicolumn{1}{|c}{$\Phi^{-}$}\\\hline
$X^{A_{1}}$ & $Z$ & \multicolumn{1}{|c}{$\Psi^{+}$}\\\hline
$X^{A}X^{A_{1}}$ & $ZX$ & \multicolumn{1}{|c}{$\Psi^{-}$}\\\hline\hline
\end{tabular}
\caption{The table details the results of Bob's Bell state
analysis. The states in the third column are syndromes that determine
the channel error (``Error'') and the recovery operation (``Recovery'') that Bob
should perform to recover the initial information photon.}\label{tbl:syndromes}%
\end{table}%

\section{Optical Encoding and Decoding Circuit}

The seminal paper of Knill, Laflamme, and Milburn (KLM)\ showed how to perform
two-qubit interactions with linear-optical devices \cite{klm,kok:135}. Their
method is a measurement-assisted scheme:\ it first mixes a set of
\textquotedblleft computational\textquotedblright\ modes and ancilla\ modes in
a linear-optical device and then counts the photons in the ancilla modes. The
optical transformation acts on the computational modes. The ancilla modes help
perform this transformation and we measure them at the end of the
measurement-assisted scheme. The gates exploit the Hong-Ou-Mandel quantum
interference of indistinguishable photons \cite{PhysRevLett.59.2044}. These
measurement-assisted transformations are heralded, non-deterministic, and
non-destructive. A destructive gate involves a measurement on the
computational modes---the informational state collapses to one of the states
in the measurement basis even when the gate succeeds. A non-destructive gate
requires a measurement only on the ancilla modes---the result is that all the
information encoded in the state remains intact when the gate succeeds
\cite{bao:170502,kok:135}.

The measurement-assisted scheme is useful for dual-rail encoded qubits and
even polarization-encoded qubits \cite{spedalieri:012334}, but until now, no
one has considered its application to \textquotedblleft
quad-rail\textquotedblright\ encoded quantum information in polarization-OAM\ states.

The implementation of a polarization-OAM measurement-assisted scheme requires
unconventional linear-optical elements analogous to beamsplitters and other
tools of linear optics acting on OAM states of photons. Holographic elements
suffice for this purpose because they act on OAM components
\cite{nat2008kwiat}---similarly to the action of polarization beamsplitters on
photon polarization. The extension of the measurement-assisted scheme to OAM
states is a generalization of the idea in Ref.~\cite{spedalieri:012334}.
There, the authors extended the measurement-assisted scheme to
polarization-encoded qubits. Here, we use a similar idea to extend the
measurement-assisted scheme to OAM states.

The encoding transformation corresponding to our code generates the encoded
state in (\ref{eq:encoded-state}). It is a controlled-sign gate that acts on
the four-dimensional Hilbert space $\mathcal{H}_{A}\otimes\mathcal{H}_{A_{1}}$
of the information photon $A$ and the polarization subspace of Alice's part
$A_{1}$ of the hyperentangled state in\ (\ref{eq:unencoded-state}). The gate
acts on the polarization degrees of freedom and leaves the OAM degrees of
freedom invariant. It is a linear-optical transformation on a set of six
modes---two modes for the $A$ system and four modes for the $A_{1}$ system.
The gate leaves the following basis states%
\begin{align}
&  \left\vert H\right\rangle ^{A}\left\vert H\circlearrowleft\right\rangle
^{A_{1}},\ \ \left\vert H\right\rangle ^{A}\left\vert H\circlearrowright
\right\rangle ^{A_{1}},\ \ \left\vert H\right\rangle ^{A}\left\vert
V\circlearrowleft\right\rangle ^{A_{1}},\nonumber\\
&  \left\vert H\right\rangle ^{A}\left\vert V\circlearrowright\right\rangle
^{A_{1}},\ \ \left\vert V\right\rangle ^{A}\left\vert H\circlearrowleft
\right\rangle ^{A_{1}},\ \ \left\vert V\right\rangle ^{A}\left\vert
H\circlearrowright\right\rangle ^{A_{1}},\label{eq:encoding-transformation1}%
\end{align}
invariant and adds a phase to the remaining basis states:%
\begin{align}
\left\vert V\right\rangle ^{A}\left\vert V\circlearrowleft\right\rangle
^{A_{1}} &  \rightarrow-\left\vert V\right\rangle ^{A}\left\vert
V\circlearrowleft\right\rangle ^{A_{1}},\nonumber\\
\left\vert V\right\rangle ^{A}\left\vert V\circlearrowright\right\rangle
^{A_{1}} &  \rightarrow-\left\vert V\right\rangle ^{A}\left\vert
V\circlearrowright\right\rangle ^{A_{1}}.\label{eq:encoding-transformation}%
\end{align}

We make a statement about the mathematical structure of the Hilbert space of
polarization-OAM\ states. It is possible to decompose any Hilbert space with a
tensor product structure. E.g.,
\begin{multline*}
\text{span}\left\{  \left\vert H\circlearrowleft\right\rangle ,\left\vert
V\circlearrowleft\right\rangle ,\left\vert H\circlearrowright\right\rangle
,\left\vert V\circlearrowright\right\rangle \right\} \\
=\text{span}\left\{  \left\vert H\right\rangle ,\left\vert V\right\rangle
\right\}  \otimes\text{span}\left\{  \left\vert \circlearrowleft\right\rangle
,\left\vert \circlearrowright\right\rangle \right\}  .
\end{multline*}
While in the paraxial approximation, local optical transformations on the
subspaces span$\left\{  \left\vert H\right\rangle ,\left\vert V\right\rangle
\right\}  $ and span$\left\{  \left\vert \circlearrowleft\right\rangle
,\left\vert \circlearrowright\right\rangle \right\}  $ respect the
tensor-product decomposition of the full four-dimensional space. However, the
tensor-product notation in (\ref{eq:ebit}) from Ref.~\cite{nat2008kwiat} may
be somewhat misleading in the context of a measurement-assisted transformation
(which the authors of Ref.~\cite{nat2008kwiat} do not discuss). A
qubit-coupling measurement-assisted transformation, based on the mixing of
creation operators in separate modes, does not respect such a tensor-product
decomposition in general. Instead such an operation acts naturally on a space
constructed as a direct sum, e.g., span$\left\{  \left\vert H\circlearrowleft
\right\rangle ,\left\vert V\circlearrowleft\right\rangle \right\}  \oplus
\ $span$\left\{  \left\vert H\circlearrowright\right\rangle ,\left\vert
V\circlearrowright\right\rangle \right\}  $. The above restriction places a
constraint on the form of the linear-optical encoding circuit and decoding circuit.

Knill devised an optimal solution for the controlled-sign gate
\cite{PhysRevA.66.052306}. We \textit{could} use Knill's two dual-rail qubit
gate for an implementation of the transformation in
(\ref{eq:encoding-transformation}). It requires the combination of two
separate transforms: the first acts on the $\circlearrowleft$ OAM states and
the second acts on the $\circlearrowright$ OAM states. Each transformation
requires two ancilla modes and has a success probability of $2/27$
\cite{PhysRevA.66.052306}. This \textquotedblleft Knill
combination\textquotedblright\ scheme thus requires four ancilla modes with a
success probability of $\left(  2/27\right)  ^{2}\approx0.0055$.

We have employed the numerical optimization technique in
Ref.~\cite{prep2008uskov} for our encoding circuit rather than the above Knill
combination scheme. Our optical scheme for the encoding transformation in
(\ref{eq:encoding-transformation1}) and (\ref{eq:encoding-transformation}%
)\ requires only three ancilla photons and has a success probability of 0.0097
(nearly a two-fold increase over the Knill combination scheme).

We briefly describe the numerical optimization technique for finding a general
linear-optical circuit \cite{prep2008uskov}. An $N\times N$ matrix, where $N$
is the number of optical modes, completely characterizes an optical
transformation. The numerical implementation of the optimization algorithm
performs a gradient search in the space of these matrices. The algorithm first
finds a matrix that guarantees a unit transformation fidelity and then
performs a constrained optimization of success probability in the unit
fidelity subspace. Each optimization cycle ends at some local maximum of
success probabilitiy. The resulting Kraus transformation, or contraction map,
acting on the computational modes matches the desired target transformation in
the case of a successful measurement outcome on the ancilla modes.%

\begin{figure}
[ptb]
\begin{center}
\includegraphics[
natheight=1.719300in,
natwidth=2.000300in,
height=2.8253in,
width=3.2776in
]%
{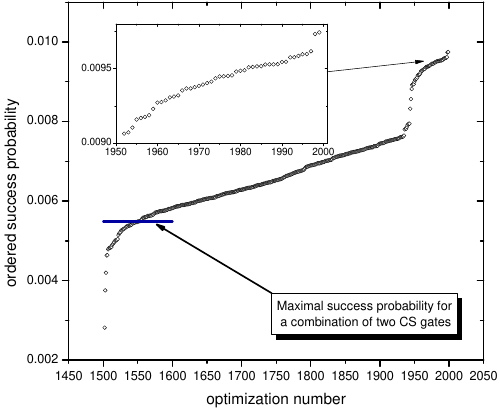}%
\caption{The figure displays the results of the numerical gate optimization
algorithm for our case of six computational modes and three ancilla modes. The
algorithm starts an optimization at a randomly selected point in the space of
$6\times6$ matrices and maximizes the gate fidelity. The second stage of the
optimization algorithm is a constrained optimization of success probability
within the unit fidelity subspace. The figure displays ordered success
probability for the best 500 optimization cycles of the total amount of 2000
optimizations for fidelity $>1-10^{-7}$. The algorithm finds an encoding
circuit with a success probability almost twice that of the Knill combination
scheme.}%
\label{fig:numerical-results}%
\end{center}
\end{figure}
We now describe the results of the procedure for our case where the encoding
circuit acts on the modes for Alice's systems $A$ and $A_{1}$ and three
ancilla modes. We performed the optimization in the 81-dimensional complex
space of a GL$(9)$ matrix (six modes for Alice's systems $A$ and $A_{1}$ and
three ancilla modes). Such a matrix admits a decomposition into a sequence of
operations where each operation corresponds to a standard optical element
\cite{PhysRevLett.73.58}. We were able to simplify the transformation even
more by using a technique from Ref.~\cite{PhysRevA.66.052306}. The optimal
transformation actually acts on three modes only: the vertical polarization of
Alice's photon $A$, the vertical polarization and $\circlearrowright$ OAM of
Alice's photon $A_{1}$, and the vertical polarization and $\circlearrowleft$
OAM of Alice's photon $A_{1}$. The reduction of the optimal solution to a
three-mode operation should make it easier to implement experimentally. The
following equations illustrate the reduced transformation:%
\begin{align*}
\left\vert n_{1}\right\rangle _{V}^{A}\left\vert n_{2}\right\rangle
_{V\circlearrowright}^{A_{1}} &  \rightarrow\left(  -1\right)  ^{f\left(
n_{1},n_{2}\right)  }\left\vert n_{1}\right\rangle _{V}^{A}\left\vert
n_{2}\right\rangle _{V\circlearrowright}^{A_{1}},\\
\left\vert n_{1}\right\rangle _{V}^{A}\left\vert n_{2}\right\rangle
_{V\circlearrowleft}^{A_{1}} &  \rightarrow\left(  -1\right)  ^{f\left(
n_{1},n_{2}\right)  }\left\vert n_{1}\right\rangle _{V}^{A}\left\vert
n_{2}\right\rangle _{V\circlearrowleft}^{A_{1}},
\end{align*}
where $n_{1}$ and $n_{2}$ are photon numbers equal to either zero or one and
the function $f$ is equal to one if $n_{1}=n_{2}=1$ and is equal to zero
otherwise. One can verify that the implementation of the above reduced
transformation is equivalent to the full transformation in
(\ref{eq:encoding-transformation}). Figure~\ref{fig:numerical-results}
illustrates a sample distribution of success probabilities in increasing order.

The best found solution provides a success probability of 0.00974276 at a
fidelity of $1-\left(  6\times10^{-8}\right)  $. This solution, most likely,
is the optimal global solution, or at least close to the global optimum. One
cannot verify the global character of a solution by numerical tools only.
However, we have found that further optimization with the current scheme does
not improve the result. The success probability shows a slight increase for
non-unit fidelity: at the level of 0.99 fidelity, we have found a solution
with success probability of 0.011. Further relaxing the unit fidelity
requirement does not lead to any significant increase of the success probability.

We now address the issue of implementing our hyperentanglement-assisted
quantum code. As described above, our encoding circuit solution corresponds to
some optical transformation matrix. The first question arising in connection
with the possible experimental implementation of such an optical
transformation is whether the obtained matrix is unitary. Currently, we use an
optimization algorithm that searches for an arbitrary matrix, one not
necessarily restricted to the subspace of unitary matrices. Therefore, one may
need to apply a unitary dilation procedure
\cite{PhysRevA.66.052306,VanMeter:2007:063808} to embed the matrix into a
larger unitary matrix. However, a solution corresponding to a maximal success
probability is a matrix requiring minimal dilation, due to the singular
behavior of the gradient of the success probability function on the manifold
of unitary matrices. Indeed the two best solutions (see insert in Figure 2)
correspond to matrices with the following singular values $\left\{
1.,1.,1.,1.,1.,0.5\right\}  $. One needs to introduce only one additional
vacuum mode to embed the matrix into an SU$(7)$ matrix in order to implement
the transformation in the form of beamsplitters and phase shifters.
Ref.~\cite{PhysRevLett.73.58} suggests the scheme of such a decomposition.
Mathematically, it corresponds to a factorization of an arbitrary SU$(N)$
matrix into a product of SU$(2)$ matrices. The implementation in
Ref.~\cite{PhysRevLett.73.58} requires $N(N-1)/2$ beam splitters. However, it
is well known that in some important cases one can find a simpler
decomposition \cite{PhysRevA.66.052306}. Knill's scheme requires only four
beamsplitters for the realization of an SU$(4)$ matrix, whereas the original
KLM scheme requires eight beam splitters for an SU$(6)$ matrix. The matrix we
have obtained is rather complicated because we were not able to find a
\textquotedblleft nice\textquotedblright\ decomposition (See the Appendix for
the matrix).

\section{Conclusion}

We have presented an optical implementation of an entanglement-assisted
quantum code that should be realizable with current technology. The code
encodes one information photon with the help of a hyperentangled state. To our
knowledge, this proposal is the first suggestion for an implementation of an
entanglement-assisted quantum code.

\section{Acknowledgements}

The authors thank Todd A. Brun for useful discussions. M.M.W. acknowledges
support from NSF Grant 0545845 and D.B.U. acknowledges support from the Army
Research Office and the Intelligence Advanced Research Projects Activity.

\appendix

\section{}

Below we list the matrix that implements the transformation for the encoding
circuit. We do not provide it as one large matrix because the individual
entries are too large, but instead provide it a few columns at a time. The
first three columns are as follows:%

\[
\left[
\begin{array}
[c]{ccc}%
-0.253936+i0.215424 & 0 & -0.0269989+i0.211134\\
0 & 1 & 0\\
0.0473299+i0.183042 & 0 & -0.136174+i0.454254\\
0 & 0 & 0\\
0.196523-i0.216478 & 0 & -0.233841+i0.184769\\
0 & 0 & 0\\
-0.33549-i0.135251 & 0 & 0.314695+i0.192451\\
0.318659+i0.380869 & 0 & 0.3053+i0.314815\\
0.277613-i0.411775 & 0 & -0.0145173+i0.484746
\end{array}
\right]  .
\]
The second three columns are as follows:%
\[
\left[
\begin{array}
[c]{ccc}%
0 & -0.249991-i0.213976 & 0\\
0 & 0 & 0\\
0 & 0.262553+i0.141112 & 0\\
\alpha & 0 & 0\\
0 & -0.159651+i0.187183 & 0\\
0 & 0 & \alpha\\
0 & -0.515822+i0.243508 & 0\\
0 & 0.220057-i0.296955 & 0\\
0 & 0.039117+i0.303606 & 0
\end{array}
\right]  ,
\]
where $\alpha=-0.611421-i0.791452$. The next two columns are as follows:%
\[
\left[
\begin{array}
[c]{cc}%
0.410744+i0.0245062 & 0.367852-i0.184455\\
0 & 0\\
-0.4806-i0.326223 & 0.298488-i0.221226\\
0 & 0\\
-0.264695-i0.0518749 & 0.153492+i0.498569\\
0 & 0\\
-0.23816-i0.143929 & -0.278242-i0.00531807\\
0.132686-i0.193403 & -0.0860369+i0.41503\\
0.382029+i0.183596 & -0.229701+i0.0475146
\end{array}
\right]  .
\]
The last column is as follows:%
\[
\left[
\begin{array}
[c]{c}%
-0.0349526+i0.229345\\
0\\
-0.0337073+i0.290301\\
0\\
0.403926-i0.202786\\
0\\
-0.337872-i0.21804\\
-0.334824-i0.268026\\
-0.164744+i0.395987
\end{array}
\right]  .
\]

\bibliographystyle{apsrev}
\bibliography{optics-grandfather}

\end{document}